\newcommand{\f}[1]{Fig.~\ref{#1}}
\newcommand{\eq}[1]{Eq.~(\ref{#1})}

\def\be{\begin{equation}}
\def\ee{\end{equation}}
\def\bea{\begin{eqnarray}}
\def\eea{\end{eqnarray}}
\def\l({\left(}
\def\r){\right)}

\documentstyle[prb,aps,multicol,graphicx,psfig]{revtex}

 \renewcommand{\narrowtext}{\begin{multicols}{2} 
\global\columnwidth20.5pc}
 \renewcommand{\widetext}{\end{multicols} \global\columnwidth42.5pc}

\begin{document}
\title{Interplay of dendritic avalanches and gradual flux penetration\\
in superconducting MgB$_2$ films}
\author{D.~V. Shantsev$^{1,2}$, P. E. Goa$^1$, F. L. Barkov$^{1,3}$,
T.~H.~Johansen$^{1,}$\cite{0},  
W. N. Kang, and S. I. Lee$^{4}$
}
\address{
$^1$Department of Physics, University of Oslo, P. O. Box 1048
Blindern, 0316 Oslo, Norway\\
$^2$A. F. Ioffe Physico-Technical Institute, Polytekhnicheskaya 26,
St.Petersburg 194021, Russia\\
$^3$Institute of Solid State Physics, Chernogolovka, Moscow Region, 142432, Russia\\
$^4$National Creative Research Initiative Center for Superconductivity, 
Department of Physics, Pohang University of 
Science and Technology, Pohang 790-784, Republic of Korea
}
\date{\today}
\maketitle

\begin{abstract}

Magneto-optical imaging was used to study a 
zero-field-cooled MgB$_2$ film at 9.6~K where in a slowly
increasing field the flux penetrates by 
abrupt formation of large dendritic structures.
Simultaneously, a gradual flux penetration takes place,
eventually covering the dendrites, and
a detailed analysis of this process is reported.
We find an anomalously high gradient of the flux density  
across a dendrite branch, and a peak value that decreases
as the applied field goes up.
This unexpected behaviour is reproduced by flux creep simulations based on
the non-local field-current relation in the perpendicular geometry. 
The simulations also provide indirect evidence that
flux dendrites are formed at an elevated local temperature, consistent
with a thermo-magnetic mechanism of the instability.     
\end{abstract}


\narrowtext

\section{Introduction}

The dendritic flux instability found in thin films of various
superconducting materials is a striking but still 
poorly understood phenomenon.  
It consists in an abrupt penetration of magnetic flux
into the superconductor along narrow branching channels
which form irregular dendritic patterns on the 
macroscopic scale.
The dendritic instability has been observed by
magneto-optical (MO) imaging in films of
Nb\cite{1967,duran,vv},  
YBa$_2$Cu$_3$O$_7$ \cite{leiderer,bolz} (induced by a laser pulse),
and recently in  
MgB$_2$\cite{epl,phc,apl,prb} and Nb$_3$Sn\cite{Nb3Sn}.

The instability is believed to be of thermomagnetic origin, 
similarly to the much more explored phenomenon of flux jumping.\cite{wipf,mints}
Local heating due to flux motion reduces the pinning, and
facilitates the further motion, which may lead to an
avalanche process accompanied by a substantial temperature rise.
This thermal mechanism behind dendrite formation is supported
by a recent experiment, which showed that the instability can be suppressed by 
having a normal metal in contact with the superconductor.\cite{phc}
Dendritic patterns of flux\cite{epl} and temperature\cite{aranson}
have also been obtained by simulations based on the thermal feedback mechanism.

From the thermo-magnetic nature of this instability, one expects 
that the critical current density $j_c$
characterizing the flux profile across the dendritic branches reflects
the elevated temperature at which they were formed.
Due to a decrease of $j_c$ with temperature, these profiles should have a less
steep slope than the profiles of the 
regular and smooth penetration from the edges.
Surprisingly, we find that in films of MgB$_2$
the flux profiles across the dendritic branches are actually much steeper.
In this work we investigate this paradox, and  study
using MO imaging the interplay between frozen flux dendrites and the gradually
advancing flux front. 

 \section{Experiment}

Films of MgB$_2$ were fabricated on Al$_2$O$_3$
substrates using pulsed laser deposition.\cite{kang}
A 300~nm thick film shaped as a square 
with dimensions 5$\times$5~mm$^2$ was selected for the present studies.
The sample has a high degree of $c$-axis alignment perpendicular to the plane,
and shows a sharp superconducting transition at $T_c=39$~K.

The flux density distribution in the superconducting film was visualized
using MO imaging based on the Faraday effect in ferrite garnet indicator films. 
For a recent review of the method, see Ref.~\onlinecite{jooss}, and a description of our 
setup is found elsewhere.\cite{joh96}
The sample was glued with GE varnish to the cold finger of the optical cryostat, and
a piece of MO indicator covering the sample area was placed loosely on top of
the MgB$_2$ film. 
The gray levels in the MO images were converted to magnetic field values
using a calibration curve obtained above $T_c$. 

The MgB$_2$ film was cooled down to different temperatures $T$
in zero magnetic field, 
and then a slowly increasing perpendicular field, $B_a$, was applied.
For $T>10$~K, the flux penetrated the film gradually, and 
formed the critical-state usually found
in superconductors with bulk pinning.
For $T<10$~K, the gradual penetration was interrupted by abrupt invasion
of dendritic flux structures. 
Shown in \f{f_mo} is a series of 
MO images visualizing the distribution of the perpendicular
field $B$ for $T=9.6$~K. 
At fields below 14.5~mT, a gradual flux penetration from the edges
took place. 
The distribution is then similar to the conventional 
critical-state picture in a perpendicular geometry:\cite{jooss}
the flux concentrates at the edges, seen as a bright contour around the sample,
and partly penetrates inwards. The central part and 
the regions near the corners remain flux-free and appear
black on the image.
The observed roughness of the flux front is
often found in MO studies of superconductors and
indicate the presence of small defects that lead to
the fan-like flux patterns. 
At 15~mT a big dendritic flux structure abruptly 
invaded the film (middle panel).
The properties of such flux dendrites are described in detail in the 
previous studies.\cite{epl,sust,phc,apl,prb}
Here we emphasize that   
the exact pattern of the dendrite is not reproducible, and thus
not related to defects.
During the subsequent field increase, the dendritic structure remains 
seemingly intact, while the flux front continues advancement, like
a moving sand dune, and
by $B_a=36$~mT, it covers the dendrite almost completely, see \f{f_mo}(bottom).

\vspace{0.4cm}

\begin{figure}
\centerline{\psfig{figure=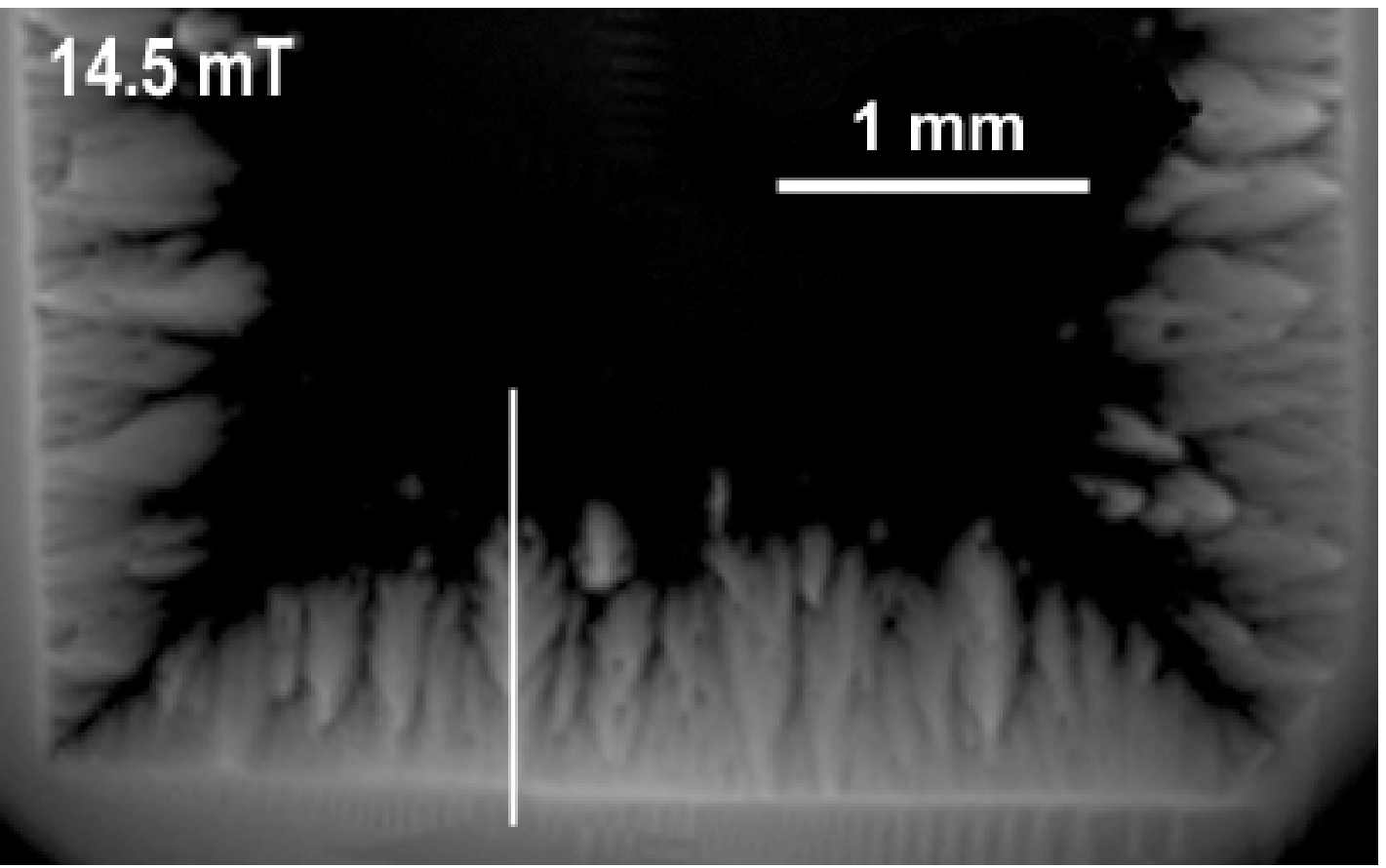,width=8.2cm}}
\centerline{\psfig{figure=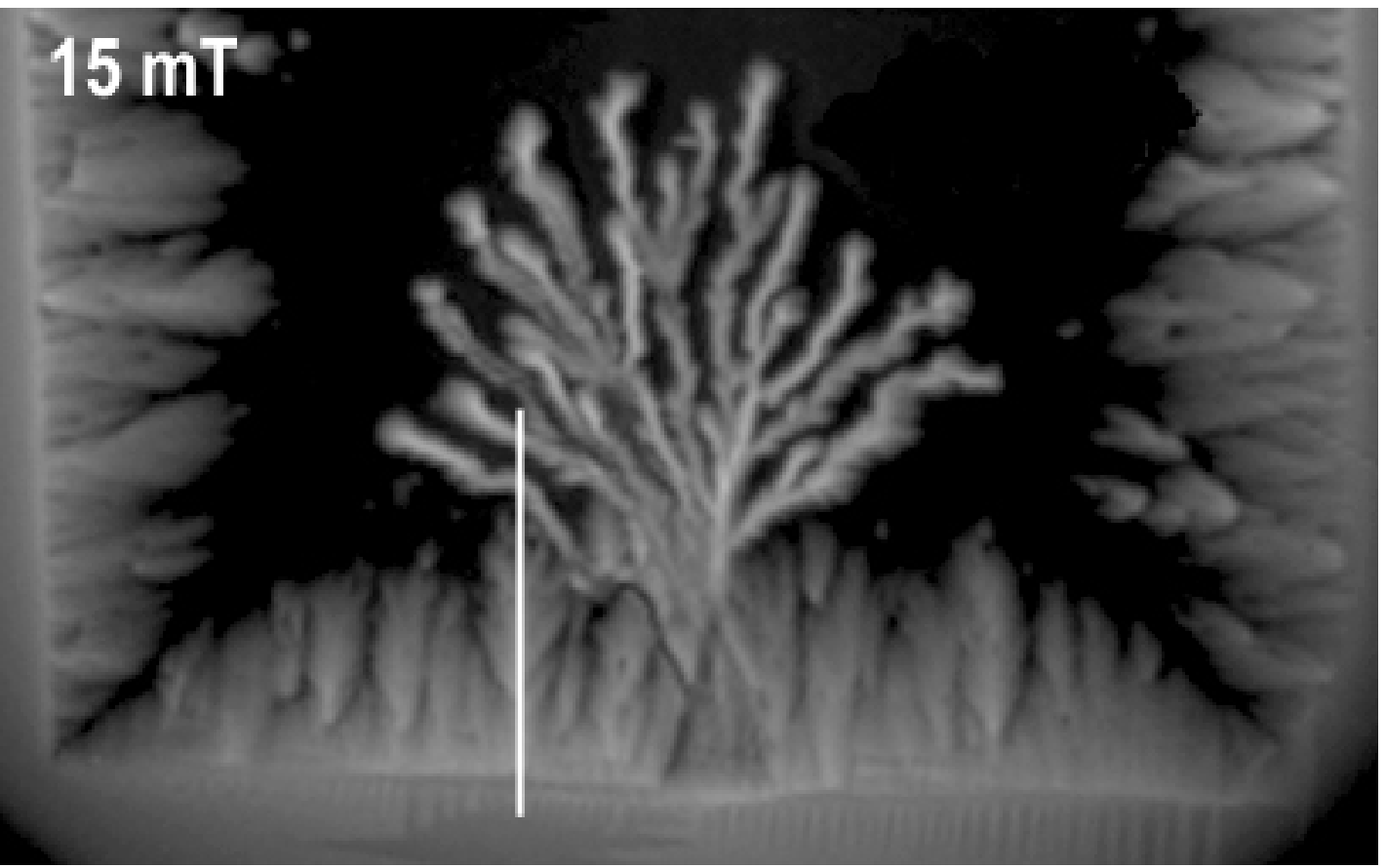,width=8.2cm}}
\centerline{\psfig{figure=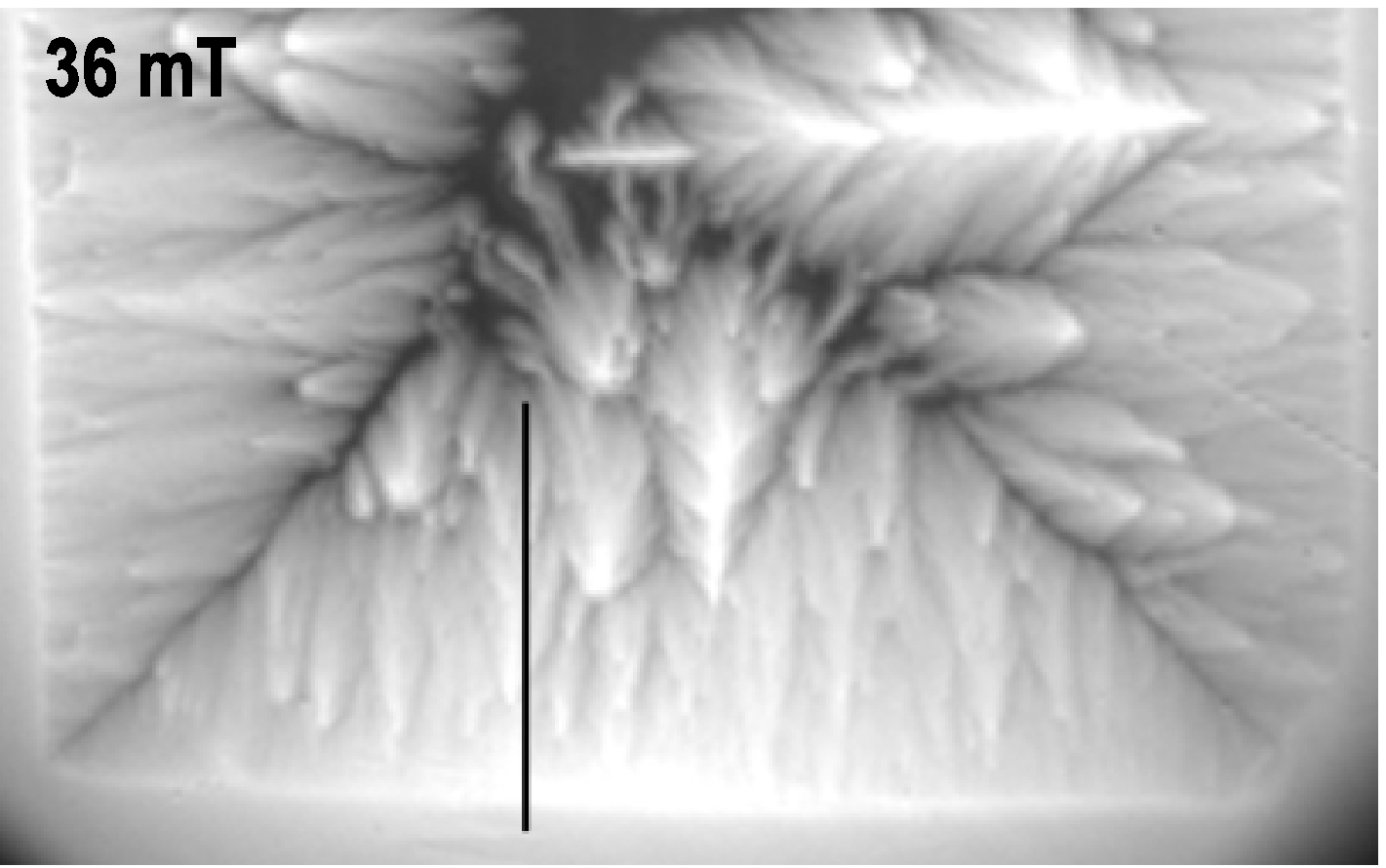,width=8.2cm}}
\vspace{0.4cm}
\caption{Magneto-optical images of flux distributions
in a MgB$_2$ film for increasing applied field 
(only the lower half of the film is seen). A dendritic structure
appeared abruptly at $B_a \approx 15$~mT.
\label{f_mo}}
\end{figure}

The evolution of the   
flux density profiles across the film during field increase
is shown in \f{f_barise}. The profiles are calculated directly from the MO
images along the line indicated in \f{f_mo}, 
which is perpendicular to both the film edge and two branches
of the dendrite tree. The slope of the profiles is thus everywhere representing
the actual $|\vec{\nabla} B|$.
The shown profiles cover all the stages
of flux penetration: the gradual advancement of flux front into the
virgin film (7 and 14.5~mT), the formation of a dendrite (15~mT), and
the later gradual penetration covering the dendrite.   
One can immediately see that  
for the dendritic branches the profile has an 
anomalously steep slope.
It is therefore tempting to conclude that this corresponds
to an anomalously high critical current density. 
However, an enhanced current density flowing around the branches is 
bewildering. High as it may be {\em during} the dendrite formation, 
the current density should relax fast, and not exceed 
$j_c$ flowing in the area of regular flux penetration.
In fact, one would expect that the heating accompanying the avalanche reduces
the current density.

Even more surprising is the observed evolution
of the flux density around a dendrite 
branch as the flux front is approaching and runs it over,
see \f{f_cut}(top).
From the Figure we again notice the anomalously steep slope of
the profile across the branch (20~mT curve) compared to 
the slope after the branch has been wiped out (33 and 39~mT curves).
In addition, one finds that the new flux coming to the region from the edge
does not simply add up to the existing $B$ distribution.
Instead, it first destroys the existing sharp peak of $B(x)$, so that 
$B$ is temporarily {\em decreasing} in the vicinity of the branch core.
These two observations are in a strong contrast
to the behavior expected in the usual Bean model,
see \f{f_cut}(bottom), where
the slopes of $B(x)$ are fixed, like in a sandpile.

\begin{figure}
\vbox{
\centerline{\psfig{figure=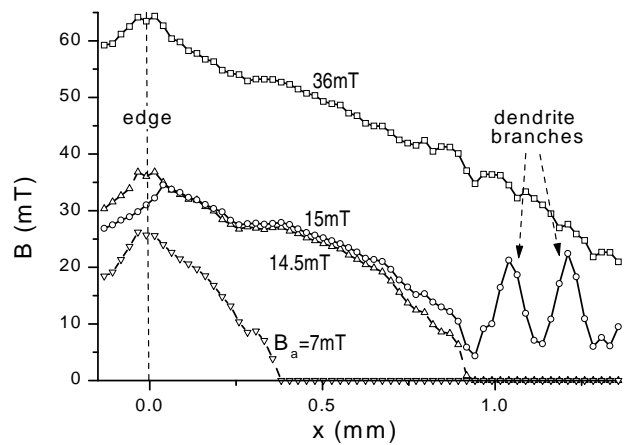,width=8.5cm}}
\vspace{0.1cm}
\caption{Profiles of flux density 
for increasing applied field obtained from MO images 
along the line shown in \f{f_mo}.
The profile slope across
the dendrite branches is much larger than that in other regions and, also
larger than in the same region for $B_a=36$~mT. 
\label{f_barise}}}
\end{figure}

\begin{figure}
\centerline{\psfig{figure=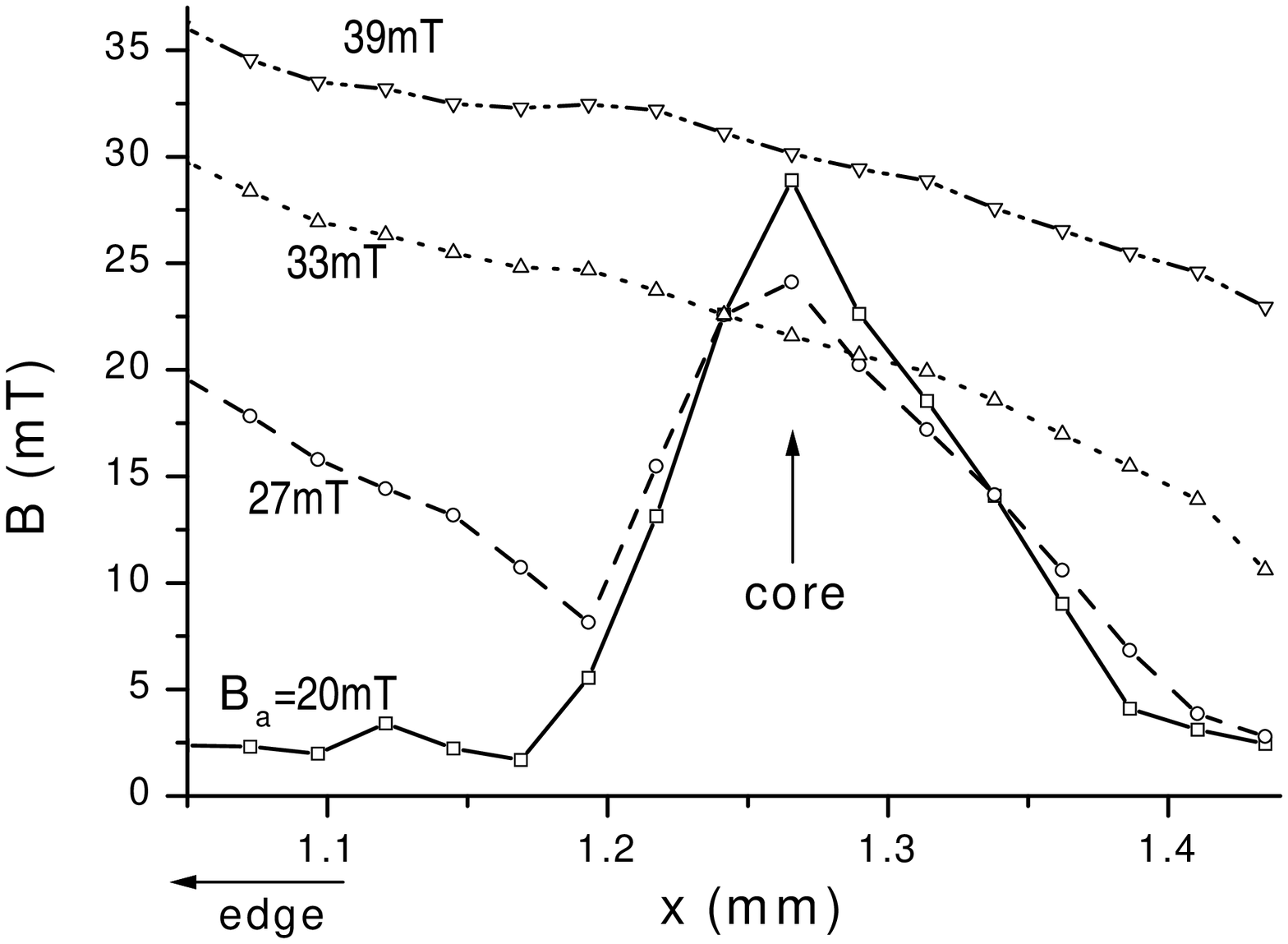,width=8.2cm}}
\centerline{\psfig{figure=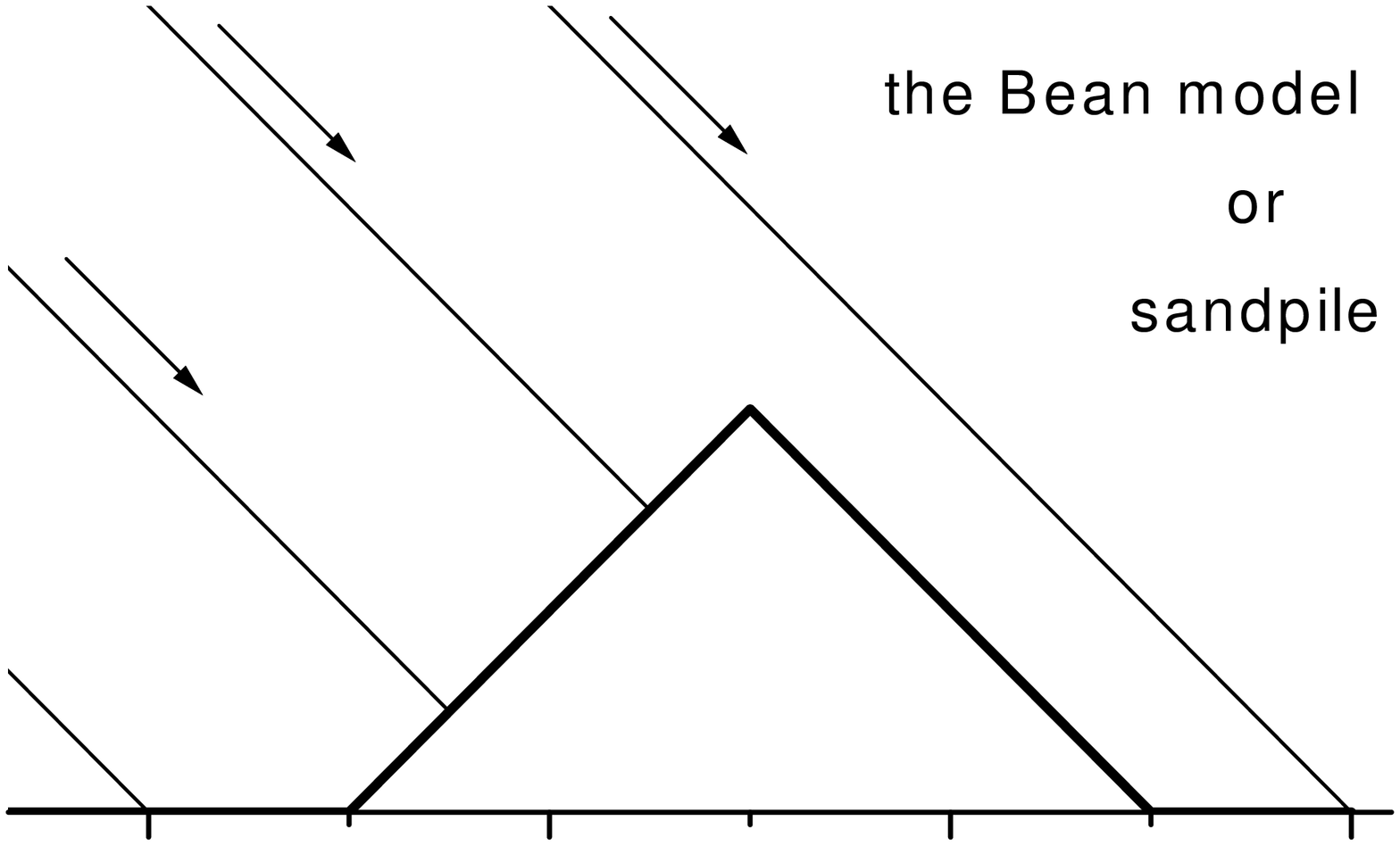,width=8.2cm}}
\vspace{0.1cm}
\caption{Top: Profiles of flux density across a dendritic branch
for increasing applied field $B_a$, where $x$ is the distance from the film edge.
The approaching flux front destroys the critical state around the branch
leading to temporal {\em decrease} of local flux density in the peak region.
Bottom: Behavior expected for a sandpile or the Bean model
in parallel geometry, which contrasts the experimental results.  
\label{f_cut}}
\end{figure}

\section{Simulations}


To seek an explanation for the surprising observations
we take into account the fact that our sample is a thin film.
In such a perpendicular geometry the relation between
$B$ and $j$ is non-local, and this might be responsible for the 
experimental behavior shown in \f{f_cut}.
We address the problem by 
calculating the evolution of $B$ and $j$ distributions for
the case of flux front approaching a dendritic branch,
using flux creep simulations.
Focus is set on the flux profile development after the dendrite 
is formed, while the dendritic branch itself is introduced 
``by hand'' at certain applied field.

Consider a thin film strip with width $2w$ along the $x$-axis, 
and thickness $d$ along the $z$-axis.
We assume $d \ll w$ and neglect variation of all quantities throughout 
the strip thickness. A magnetic field applied perpendicular to the strip
will then induce a sheet current $J=dj$, and an electric 
field $E$ directed along the $y$-axis. From the Maxwell equation,
one has 
\begin{eqnarray}
\partial B / \partial t &=& -  \partial E /\partial x \, .
\label{dBdt}
\end{eqnarray}
Superconductors in the flux creep regime are usually
described by the current-voltage relation,
\be
    E(J,B) = v_0 |B| \; \left| J/J_c\right|^n  {\rm sgn} J \, ,
\label{EBJ}
\ee
where $n \gg 1$, and has the meaning of a vortex depinning activation 
energy divided by $kT$,\cite{brandt-rev} and $v_0$ is
the vortex velocity at $J=J_c$.
Finally, the nonlocal relation linking 
the current and flux density distributions in the strip reads\cite{BrIn} 
\be
\mu_0 J(x) = \frac{2}{\pi} \int^{2w}_{0} \frac{B(x')-B_a}{x-x'}
\sqrt{\frac{w^2-(x'-w)^2}{w^2-(x-w)^2}} \ dx' \, .
\label{JB}
\ee  

The simulations start with zero initial conditions, 
$B(x,0)=E(x,0)=J(x,0)=0$,  
and for $t>0$ the applied magnetic field is assumed linearly 
increasing in time, $B_a=\dot B_a\ t$. 
To provide correspondence with the experiment,
the field ramp rate was very slow: $\dot B_a \ll B_c v_0/w $, 
where $B_c=\mu_0 J_c/\pi$ is a typical values for the flux density.
The evolution of flux and current density distributions 
$J(x,t)$, $B(x,t)$ is then calculated numerically from
Eqs.~(\ref{dBdt})-(\ref{JB}).
At the field $B_a=B_a^*$ a dendritic branch parallel to the film edge
is introduced by setting 
\be 
B(x)=\alpha B_a^*, \quad x_0 \le x \le x_0+\Delta x \, ,
\label{dend} 
\ee
where $x_0$ and $\Delta$ give the location and width of the branch.
Physically, this is equivalent to having the region
$x_0 \le x \le x_0+\Delta x$
instantly heated to the normal state so that it becomes uniformly
penetrated by flux with a density proportional to the applied 
field $B_a^*$, where the coefficient $\alpha$ 
is determined by the demagnetization factor. The simulations then 
continue
with the same parameters, i.e., we assume that the heated region 
cools down immediately. To mimic our actual experimental situation,
seen in \f{f_barise}, the following parameter values were chosen: 
$x_0/w=0.5$, i.e., the dendrite is formed
halfway to the film center, at the applied field $B_a^*=0.58B_c$ 
when the flux front is located at $\approx 0.75 x_0$, and to have
appropriate height and width of the peak at $x_0$ we set
$\alpha=2$ and $\Delta/w=0.03$. 

Let us first analyze the distributions for the right half 
of the strip, where almost no disturbance is created by 
the dendrite, see \f{f_prg}. These profiles demonstrate
a familiar scenario of flux penetration in the perpendicular geometry: 
nonlinear $B(x)$ with a flux front advancing deeper and deeper as $B_a$
increases, and essentially uniform $J(x)$ in the flux penetrated area 
and a considerable current also in the Meissner state central part.\cite{endnote}
Such profiles are often found in MO studies of thin superconductors,\cite{joh96}
and were obtained by flux creep simulations already long ago.\cite{schuster}    
The profiles are also very close to the Bean-model
result for a thin strip,\cite{BrIn,zeld} which is expected for
the large $n=25$ used in the present simulations.

Next, we examine the left side of the strip, and how the 
penetration there 
is perturbed after the appearance of the dendrite. 
The initial rectangular profile, \eq{dend}, at $B_a^*$ 
relaxes very fast because of the high current density 
associated with such an artificial $B(x)$.
After this relaxation one finds a 
triangular shape of the $B$ profile across the dendrite -- see 
the thick line in \f{f_prg}.
Note that the left slope of the dendrite is steeper than on the right side,
as found also experimentally, see \f{f_cut}.
As the field continues to increase the conventional penetration
advances.
It affects drastically the profile around the dendrite.
By $B_a=0.75B_c$ (dashed line), when the flux front has not yet reached the dendrite,
the peak around $x_0$ is already significantly suppressed.   
As the field reaches $0.92B_c$,  
any trace of the dendrite has disappeared, and the flux density 
at $x_0$ is only half of its original magnitude.
Besides, the slope of $B(x)$ around $x_0$ 
became much smaller than for the triangular profile.

\widetext

\begin{figure}
\vbox{
\centerline{\psfig{figure=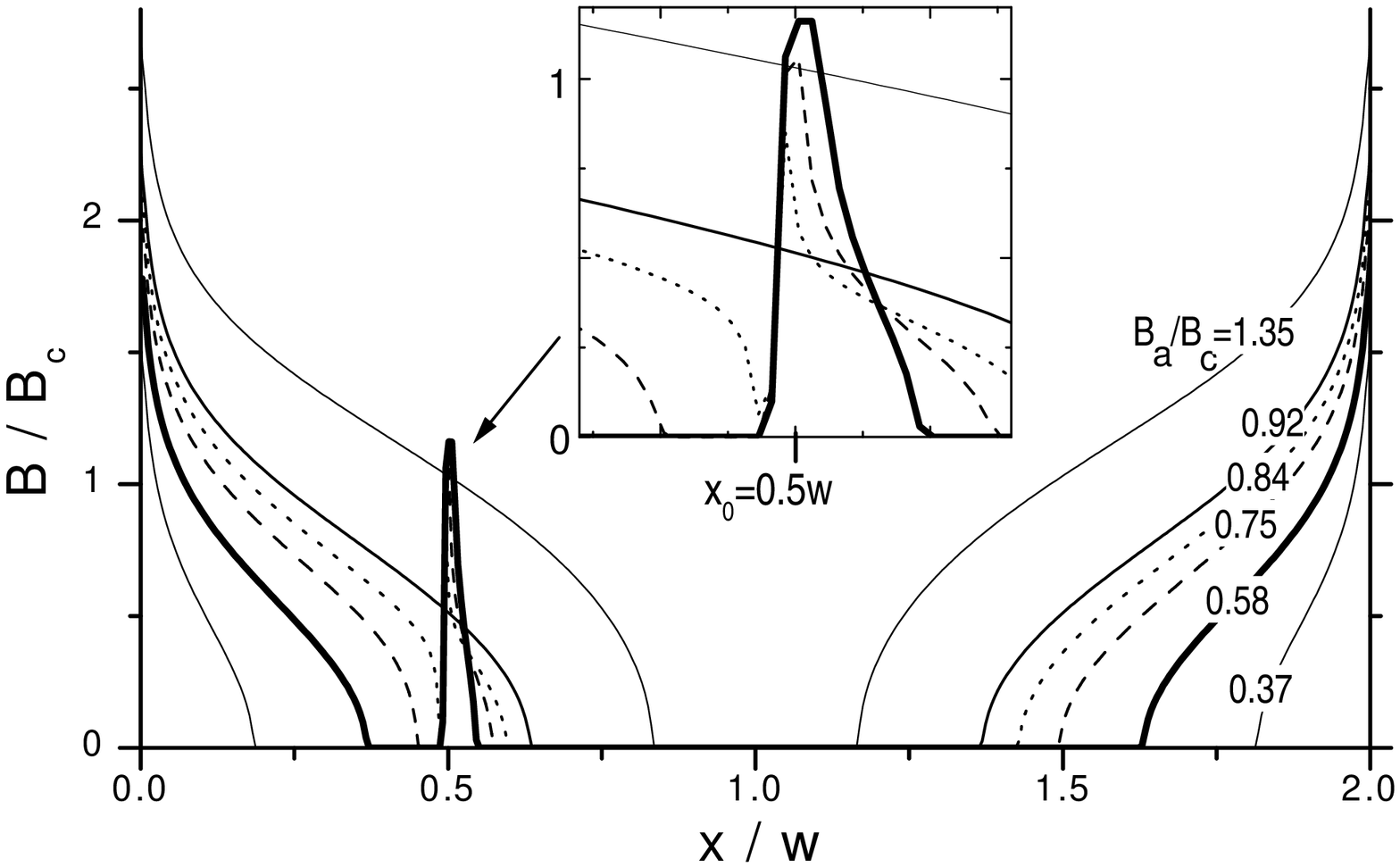,width=10cm}
\psfig{figure=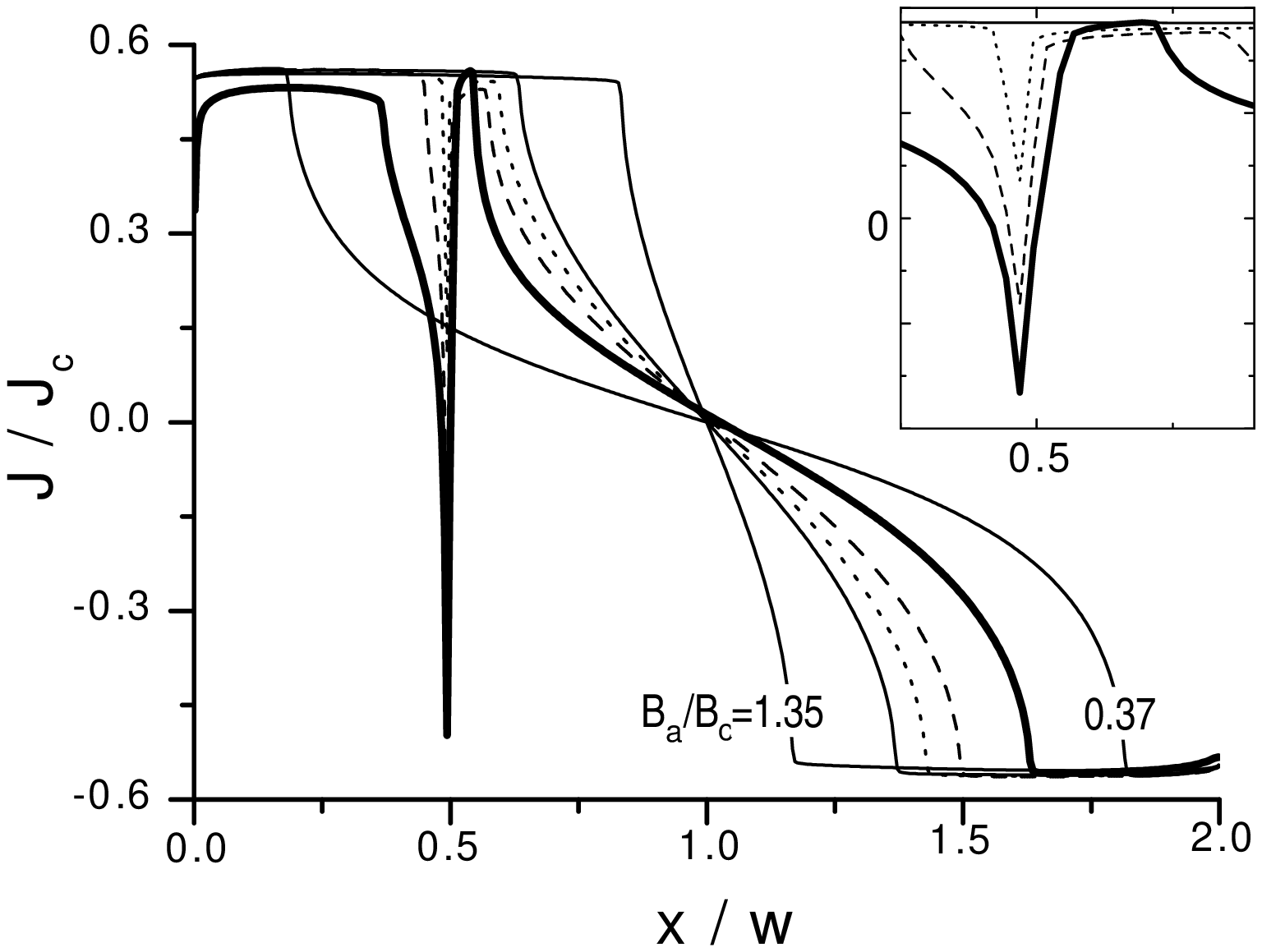,width=8cm}}
\vspace{0.1cm}
\caption{Flux creep simulations of the evolution of flux and current density
in a thin strip under increasing applied field.
At the applied field $B_a^*=0.58B_c$ a dendritic branch was
introduced at $x_0=0.5w$   
as described in \eq{dend}.
The $B$ profiles are in qualitative agreement with the experimental results shown in 
\f{f_barise} and \f{f_cut}. The insets show blow-ups of the area near the dendrite.
\label{f_prg}}}
\end{figure}
\narrowtext
  
\section{Discussion and Summary}

The present simulations clearly reproduce all main aspects of 
the experimental behaviour:\\
(i) the slope of $B(x)$ across the dendrite is anomalously steep;\\
(ii) the local flux density in the dendrite core temporarily decreases when
the flux front approaches;\\
(iii) the asymmetry of the triangular profile across the dendrite.\\
The key to understanding this behaviour lies
in the specific $B-J$ relation 
for thin strips expressed by \eq{JB}. 
This relation implies an infinite $\nabla B$
where $J$ changes abruptly, e.g.,  
near the edge and at the flux front, see \f{f_prg}. 
A similar situation is present near the dendrite, where
the current changes direction abruptly.
Consequently, $B(x)$ has anomalously high gradient there too, 
but this is in no way related to having a high local $j_c$.

Similarly, the non-locality of \eq{JB} is responsible for smearing out 
the $B$ peak at the dendrite.
Increasing $B_a$
induces additional Meissner currents $J_M$ throughout the whole strip, 
as schematically illustrated in \f{f_jtree}.
As a result, flux motion is activated on the side of the dendrite where
the current density was already $j_c$ (the side most distant from the flux front).
The flux motion proceeds in the direction of Lorentz force tending
to flatten this side of the peak. 
Earlier MO experiments have revealed similar
effects induced by 
screening currents flowing in the Meissner state region of thin films:
unexpected flux dynamics around holes\cite{holes} and slits\cite{slit}
ahead of the advancing flux front. 

\begin{figure}
\centerline{\psfig{figure=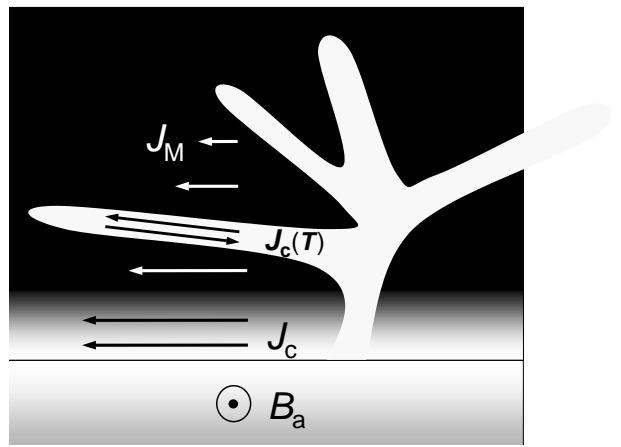,width=7cm}}
\vspace{0.2cm}
\caption{Schematic of the flux distribution and current flow in a thin film
where a flux dendrite is present.
\label{f_jtree}}
\end{figure}

If the critical state around the dendrite were formed at an elevated
temperature, the flowing current $j_c(T)$ would be correspondingly smaller.
As a result, the smearing out of the peak will start at a later stage, i.e.,
when the flux front is very close to the dendrite.
The magnitude of smearing will then also be smaller.
In fact, this is what we find by comparing
the experiments and simulations.
From the \f{f_prg} and \f{f_cut} we see that
the local decrease of $B$ at the dendrite core is
$\approx 50\%$ in the simulations, but only 
$\approx 20\%$ in the experiment.
This deviation we believe results from a local heating in the dendrite area
during its formation, an effect which was ignored in the simulations.

The work was financially supported by the Norwegian 
Research Council, NorFa and FUNMAT/UiO.

\widetext

\begin{references}
\bibitem{endnote}
Note from the $J(x)$ profiles in \f{f_prg}
that the apparent critical current density is
$\approx 0.6 J_c$, which is determined by the 
characteristic electric field $E=w\dot B_a$ and \eq{EBJ}.
\end{references}
\end{document}